\documentclass[11pt,letterpaper]{llncs}
\usepackage{amsmath}
\usepackage{amssymb}
\usepackage{graphicx}
\usepackage{marvosym}
\usepackage{url}
\usepackage{fancyhdr}

\newcommand{\cooo}{CO\textsubscript{2}}

\usepackage{geometry}
\newcommand{\keywords}[1]{\par\addvspace\baselineskip
\noindent\keywordname\enspace\ignorespaces#1}

\fancypagestyle{firstpage}{\fancyhf{}
\setcounter{page}{1}
\rfoot{\thepage}
}

\begin{document}


\title{\LARGE{Towards a low carbon proof-of-work~blockchain}\thanks{An earlier version of
this paper was published in David C. Wyld et al. (Eds): DMML, CSITEC, NLPI, BDBS - 2024
pp. 89-105, 2024. CS \& IT - CSCP 2024, DOI: 10.5121/csit.2024.140607}}


%
%
\author{\large{Agron Gemajli \and Shivam Patel \and Phillip G. Bradford}}
\institute{\large{School of Computing, University of Connecticut, USA}}


\maketitle

\thispagestyle{firstpage}

\begin{abstract}
Proof of Work (PoW) blockchains burn a lot of energy.
 Proof-of-work algorithms are expensive by design and often only serve to compute blockchains.
 In some sense, carbon-based and non-carbon based
 regional electric power is fungible. So the total carbon and non-carbon electric power mix plays 
 a role.
 Thus, generally PoW algorithms have large \cooo \ footprints solely for computing blockchains.
A proof of technology is described towards replacing hashcash or other PoW methods with a lottery and 
proof-of-VM {\em (PoVM)} emulation.
 PoVM emulation is a form of PoW where 
 an autonomous blockchain miner gets a lottery ticket in exchange for providing a {\em VM (virtual Machine)} 
 for a specified period.
 These VMs get their jobs from a job queue.  
 Managing and ensuring, by concensus, that autonomous PoVMs are properly configured and running as expected 
 gives several gaps for a complete practical system. 
 These gaps are discussed.
 Our system is similar to a number of other blockchain systems. 
 We briefly survey these systems.
 This paper along with our proof of technology was done as a senior design project.

\keywords{Blockchain, low carbon blockchain}
\end{abstract}


%
%
\section{Introduction}

This proof of technology assumes some familiarity with blockchains such as can be found in~\cite{chains-that-bind-us}.

{\em Proof-of-work (PoW)} blockchains construct hash chains that (1) have subchains that seem to be 
intractable to counterfeit
 and (2) their proofs-of-work are designed to throttle the network use and abuse while allowing select 
 workers (miners) to get paid.
 These characteristics are central for building trust for autonomous actors to use, create, and maintain 
 PoW blockchains.
 
Proof-of-work blockchains consume a lot of energy.
For instance, the bitcoin network uses as much power as small countries~\cite{NYTimes2021}. 
Consider current regional energy supply mixes and the transportable nature of blockchain mining.
These energy supply mixes and regional fungibility of electricity
 give way to blockchains having large \cooo \ footprints.

{\em Proof-of-stake (PoS)} blockchains require their participants to put up assets as collateral. 
Usually, this collateral is in cryptocurrency.
These participants validate the blocks in the blockchain.
This validation is akin to building the blocks in a PoW blockchain.
A winner (who gets paid) is randomly selected from the participants who put up collateral.
This may be done by a multiparty lottery.
This requires less resources and thus produces less \cooo.
Our proof of technology is different since we target PoW blockchains.
It is notable that many recent systems that can offer low \cooo \ are either PoS or use 
variations of PoW and PoS.

This prototype low \cooo \ PoVM system uses
 Docker\textsuperscript{\textregistered} containers orchestrated by independent instances of {\em Kubernetes (K8s)}.
 Each instance of K8s manages containers to act as VMs for the PoVMs.
 These PoVMs can be arranged as complex networks of containers by K8s.
 Skupper manages the K8s instances and sends jobs to the K8s instances. 
 These K8s instances then run the jobs on the Docker containers.
 Replace a PoW blockchain with a PoVM so it may do useful work {\em while} building the blockchain.
 This useful work is submitted by customers to the job queue.
 These jobs are executed by the Docker PoVMs.
 The work done by these VMs becomes a proxy for PoW.
 
 Ideally, these customer jobs would be done even if the blockchain did not exist.
 Thus, this system has the potential to lower the \cooo \ footprint.

The concept of low \cooo \ distributed peer-to-peer computing began with volunteer distributed computing
systems.
It has evolved to scientific computing problem instances
as PoW problems.  In several of these systems there is a mix of PoW and PoS.
We describe more details of these systems in the previous work subsection.
The focus of the current paper is towards PoVM systems using off-the-shelf systems in a senior-design project.
This senior-design project took a total of 6-credit hours split over two semesters.

This paper describes the start of a PoVM blockchain targeting a lower \cooo \ footprint than 
traditional PoW blockchains. 
This will be accomplished by replacing PoW blockchains with PoVM blockchains.
This is assuming the source of power is (mostly) carbon based.
The VMs we focus on are OS-level VMs or containers.
OS-level VMs are lightweight VMs.
Particularly, OS-level VMs share infrastructure of their host machines.
This keeps the cost of these VMs low. 
This shared infrastructure may also subject these containers, or their hosts,  to security weaknesses.

Intuitively, a PoVM supplies a general computation plaform.
This platform should be capable of running in any number of standard environments.
These include any images available on Docker or any image that can be built on Docker.
These general computational platforms run programs and their data is supplied by customers.
These customers will supply computing job instances to our system.

\begin{definition}[Jobs, transactions, and memcache]
{\sf
A job is a basic computational unit of value that produces
a single answer.
A job is computable in a single OS-level VM container.

A transaction is a token or currency transaction performed in a blockchain.
The memcache is a queue of transactions that
will be memorialized in the blocks of a blockchain.
}
\end{definition}

A job can be a decision problem so it only has either a {\em true} or {\em false} answer.
Though jobs need not only be decision problems.

A central challange is to prove these jobs were completed properly with the appropriate resources.
Of course, a customer may supply a problem instance that cannot be completed within the
given resources. 
This is where service-level agreements come in.

\begin{definition}[Service-Level Agreement (SLA)]
{\sf
An SLA is a pre-agreed set of terms and conditions about a supplied VM and jobs it may run.
}
\label{SLA}
\end{definition}

An SLA may be two-sided. That is,
the PoVMs may have to abide by certain limits. 
Also the customers' instances may have to abide by certain limits.

In our context, an SLA includes the number of instructions-per-second, flops, the amount of available core
memory, the amount of persistant storage, speed of the bus and other subsystems.
Ensuring a PoVM adheres to an SLA may be reputation based.
Several systems use reputation as a proxy for validating adherence to expectations.
Likewise, customers may get reputation rankings for supplying problem instances that 
can be solved within the agreed or negotiated SLAs.

It is notable that dockerfiles can structure system configurations.
Similarly, Kubernetes configuration files are a start towards SLAs.
These configuration files lack certain useful terms such as
instructions-per-second.
Indeed, K9s configuration files may cover SLAs for container orchestration, but not other
issues for running jobs.

Just like PoW blockchains, PoVM blockchains have subchains that are 
(1) apparently intractable to counterfeit
 and (2) the proof-of-VM is designed to throttle the network use and abuse.
  These VMs must be managed and validated by concensus.

\begin{definition}[Proof-of-Virtual-Machine (PoVM)]
{\sf
Consider a VM $M$, an SLA $S$, a configuration $C$, and a job $J$.

A PoVM is a VM $M$ and a publically verifiable evidence that $M$ has configuration $C$
and $M$ runs $J$ for a fixed period of time while satisfying the SLA $S$.
}
\label{PoVM}
\end{definition}

Showing a computer ran a particular job is challenging. This is is particularly
hard for jobs that only have a Boolean output.
However, there are a number of techniques that can help.
Retaining logs and checkpointing can both supply evidence for a PoVM.
Checkpointing saves program data and state at different stages. 
Checkpointing is traditionally a fault-tolerant mechanism though we are not interested in checkpoints for fault tolerance.

There are several technical gaps to fill-in to make our proof-of-tech useful.
For example, verifying adherence to SLAs,
validating a properly configured system, and validating that a VM is continuously working 
according to the SLA.
Several systems, for example~\cite{golem,iExec}, use reputation-based metrics to 
evaluate likelihood of compliance going forward as well as the behavior of customers.

Smart contracts are usually computed by all miners in blockchain networks. 
In other words, smart contracts are computed {\em on-chain}.
Thus, on-chain PoW computation generally produces a lot of \cooo.
The PoVMs are not on-chain, though records of their verification are on-chain.
The same holds true for the records of the PoW calculations. 
That is, PoW verifications are stored in blocks but PoW calculations are not done on-chain.
PoVMs are designed to run standard loads as PoW proxies.
Neither PoW nor PoVMs are run on-chain.

It is in the egalitarian spirit of many PoW blockchains to allow anyone to
offer their mining services.
So, the PoVM proposal given here can be enhanced by homomorphic encryption.
This is to prevent an organization's  competitors, or other interested parties, from
accessing their computations.
Homomorphic encryption is getting cheaper and faster~\cite{G,GSW,BV,BGV,CGGI}.
Also the proposal in this paper can be implemented by cloud providers themselves.
Indeed, several cloud providers offer blockchains.
In many cases, cloud users already trust their cloud service providers.
These users expect the cloud providers will keep their systems and data secure.
Homomorphic encryption can help here, though it must be very efficient so the PoVMs have low \cooo \ consumption.

Khazzaka~\cite{Khazzaka} argues that, all told, the current traditional global payments 
systems require more energy than Bitcoin and Bitcoin related systems.
This argument is made by an analysis of traditional payment systems.
Their analysis includes bank transfers, credit checks, running physical bank branches, software clients and
servers, ATM machines, etc.
The energy costs they analyze also include creating and distributing new notes and coins.
{\em less} expensive.

\subsection{Hashcash PoW}
Hashcash is used for PoW blockchains.
Hashcash uses message digest hash functions to find new blocks for many PoW blockchains.
The computation using these message digest hash functions is interesting.
In the case of hashcash, these computations have no apparent application
besides restricting participation in building blocks.

A message-digest hash function is a function that maps large strings
to small fingerprints.
A fingerprint is a small string that may uniquely represent the large string input.
This representation is done by mapping a domain of large strings to a range of small strings by
a message-digest function.
Of course, by the pigeonhole principle, a set of all large strings cannot uniquely map
to a set of all small strings.
Particularly, consider the set $S$ of all strings of length 1,024-bits.
This set cannot be uniquely mapped to the set $F$ of all 8-bit strings.
Any such mapping must map many strings from $S$ to the same element of $F$ and so on.
However, ideal message-digest hash functions, which may not exist, would have the property that
if $|F|$ is large enough then it is very hard to find collisions of elements of $S$.

Example small fingerprint may be 180-bits, 256-bits or even 512-bits.
Ideally message-digest functions should run fast, but ideally it should be  intractable to find a collision 
from the message-digest of two different inputs.

Suppose $h$ is a message-digest hash function.
Given a string $B$, it appears to be intractable to find another
string $B'$ so that $h(B) = h(B')$.
This property is collision resistance.

Consider two strings $S$ and $T$. Then $ST$ is the concatenation of $S$ and $T$.

A challange that seems easier than collision resistance, is to find another string $N$ so that
$h(BN) < T$.
For example, say any $N$ will do if the message-digest hash function $h$ outputs 180-bit fingerprints.
That is, if $T = 2^{180}-1$, then $h(BN) < T$ is always true.
If $T = 2^{180}-2$ is often true for many strings $N$, etc.

Miners do their work by modifying their own nonce $N$.
Each miner can independently change their nonce.
The first miner credited with solving the hashcash proof of work gets paid.
Of course, without paying for support, blockchains would not last long. 
So, some form of payment to the miners is important.

\begin{definition}[Hashcash Proof of Work~\cite{B,DN}]
{\sf
Suppose $h$ is a message-digest hash function.
Consider a publically shared block $B$ and an individual miner-generated nonce $N$.
The first recognized miner to sufficiently modify their own nonce $N$ to generate the proscribed number of 0s 
in the initial part of the message-digest hash-function of $BN$ is declared the winner.
}
\end{definition}

Consider an ideal message-digest hash function $h$.
Computing a message-digest hash function $h$ on an input $B$ is easy.
That is, computing $h(B)$ is easy.
Finding which nonces $N$ are such that $h(BN) < T$ for some threshold value $T$ 
appears hard, but seems much easier than finding a collision.
Finding a threshold value $T$ so that $h(BN) < T$ is the same as 
finding a certain number of 0s 
in the initial part of the message-digest hash-function of $BN$.
The smaller the threshold $T$ the harder it is to find a nonce $N$ so that $h(BN) < T$.

Dwork and Naor~\cite{DN} introduced proof-of-work.
Back introduced blockchains and indicated~\cite{B}: 
``Hashcash was originally proposed as a mechanism to throttle systematic 
abuse of un-metered internet resource.''

Hashcash proof of work also contends with apparent ties for the first miner to solve the Hashcash problem.
If two or more miners are credited with winning a block simultaneously,
then declaring a winner is defered to the next mining round.
Each of the miners, simultaneously credited with winning, has its own subchain.
Each of these subchains is a potential winning subchain.
In this next mining round, miners work off of any potentially winning subchain.
Whichever of these potentially winning subchains is extended and validated first
becomes part of the blockchain. 
The other potential subchains are often abandoned.

\begin{figure}[ht]
 \begin{center}
\includegraphics[height=3cm,width=12cm]{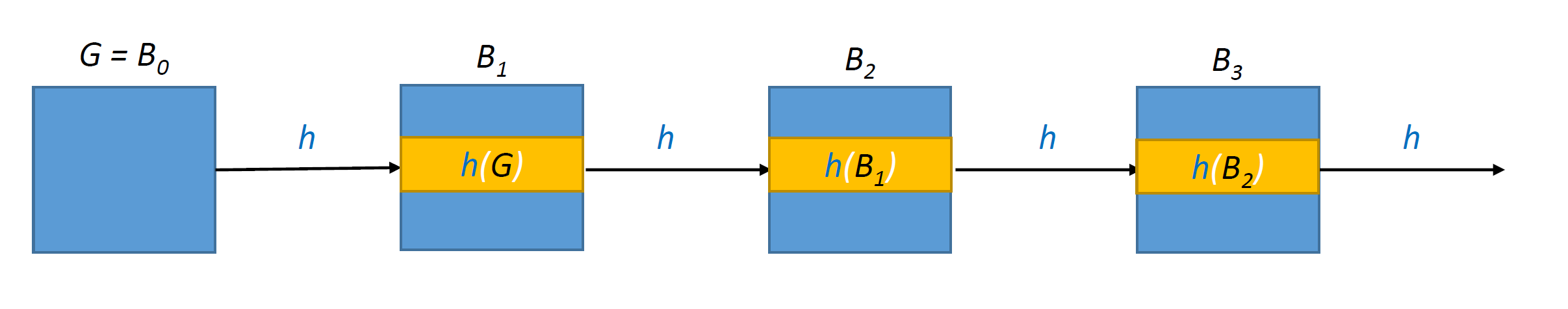}
 \end{center}
 \vspace{-0.7cm}
\caption{A blockchain-like hash chain}
\label{hashchain}
\end{figure}

Figure~\ref{hashchain} shows a tiny blockchain. 
After the genesis block $G = B_0$, the message digest hash function $h$
computes a fingerprint of the previous block concatenated with one of the miner's nonces. 
Before choosing to use a blockchain, users assume the genesis block is the valid starting block.
They also assume ideal message digest hash functions.
This assumes: it is intractable to find two distinct 
inputs that have the same fingerprint. 
Thus, supposing an ideal message-digest hash function $h$ and a single input block $B$.
Two miners can generate a different inputs $BN_1$ and $BN_2$ based 
on their own independent choices of nonce $N_1$ and $N_2$.
It seems intractable to find  two different nonces $N_1 \neq N_2$ so that $h(BN_1)= h(BN_2)$.
So, traditional PoW blockchains leverage the Hashcash-PoW challenge of computing the same fingerprint 
with different inputs ($BN_1$ and $BN_2$).
In both cases, each miner has no apparent advantage in finding a suitable modified input
$BN_1$ and $BN_2$ to ensure $h(BN_1)$ or 
$h(BN_2)$ are sufficiently small.
Hashcash-PoW seems to depend on effort and luck.

We propose replacing Hashcash-PoW with a lottery along with PoVM.
 The winner of the lottery gets paid, in essence, 
for supporting the network.

The number of lottery tickets is proportional to the number of PoVMs a miner successfully emulates.
This emulation is assumed to be with the given configuration and the given SLA.
These VMs run jobs from the job queue. 
The cost of VM emulation will throttle miners from abusing the PoVM system.
The blocks contain transactions from the mempool. These
blocks are joined by message digest hash functions, but without
doing Hashcash-PoW computations.

In the start of this proof of technology, we 
 use Docker,  Kubernetes (K8s), and Skupper.
K8s manages and monitors container loads. It also uses heartbeats to monitor containers.
Skupper connects K8s instances.
We have not implemented the configurations or any SLAs.

\subsection{Previous work}

Voluntarily sharing idle computer time has been around since at least the mid 1990s~\cite{gimp}.
See a few other early volunteer systems in~\cite{folding,ChainFaaS}.
Later prominent examples include~\cite{seti,folding,boinc}.
Voluntary computing systems based on the blockchain were proposed in Shan~\cite{volunteerblockchain}.
Shan focuses on using blockchains to avoid a single-point of failure found in centeralized systems.
In addition, Shan points out the peer-to-peer nature of blockchains can enhance auditing of volunteer computing.
Adding blockchains to solve computational portions of science problems as proxys for proof-of-work 
gave way to {\em Decentralized Science (DeSci)}~\cite{ss,w}.
Indeed, the DeSci problem instances are the proofs-of-work.
The cryptocurrency generated can also help sustain volunteer networks for computing DeSci challanges.
DeSci offers the potential for verification/reproducibility, trust, intellectual property tracking, 
funding, among other advantages~\cite{ss,w}.
Of course, these science problem instances as PoW proxies lower the overall \cooo \ produced.
There are several blockchain systems that serve processing power, as proofs-of-work, for general computation~\cite{Kondru,UN}.
Such real-world systems include Golem~\cite{golem}, Gridcoin~\cite{Gridcoin}, 
iExec~\cite{iExec}, and SONM~\cite{SONM}.
These blockchains are a mix between PoW and PoS.
Gridcoin is focused on computation for science and builds on~\cite{boinc,folding}.
It is notable that Golem has recently performed substantial chemical calculations simulating 
the beginnings of life, see
Roszak, et al.~\cite{Roszak} and also the report in~\cite{heilweil}.
Golem depends on its compute suppliers to run KVM on their machines.

Ghaemi, Khazaei, and Musilek~\cite{ChainFaaS} give a serverless blockchain compute platform 
using idle computer capacity.
They use Hyperledger Fabric and a concensus system rather than PoW or PoS.

Ball, Rosen, Sabin, and Vasudevan~\cite{BRSV_2018} suggest worst-case problem instances as proof-of-work problems.
There are practical problems that have been expressed as instances of SAT.
The {\em Satisfiability Problem (SAT)} is a classical ${\cal NP}$-complete problem
The SAT appears hard in the worst-case. 
Although, it is not known how challenging solving SAT instances is on average~\cite{BT}.
Ball, et al.~\cite{BRSV_2018} highlight the assumption that the Hashcash PoW seems to provide good proofs-of-work.
However, they mention they are not aware that the Hashcash PoW problem captures any complexity class. 
So breaking Hashcash PoW problem
may only break Bitcoin and other similar systems, rather than providing a complexity theory breakthrough.
Their work is based on their earlier work: Ball, Rosen, Sabin, Vasudevan~\cite{BRSV}.
This earlier work is to solve certain graph problems that can be expressed in first order logic.

Other prior work includes, time-lock puzzles as proofs-of-work by
Bitansky, Goldwasser, Jain, Paneth, Vaikuntanathan, and Waters\cite{BGJPVW}; 
Time-lock puzzles may help with certain types of SLAs.

Philippopoulos, Ricottone, and Oliver~\cite{PRO}
and Oliver, Ricottone, and Philippopoulos~\cite{ORP} focus on ${\cal NP}$-complete problem instances.
Loe and Quaglia~\cite{LQ} also propose using ${\cal NP}$-complete problem instances as proofs-of-work.
Particularly, they offer a type of traveling salesperson problem.
They also  give a table of alternatives to Hashcash for proof-of-work.

Chatterjee, Goharshady, and Pourdamghani~\cite{CGP} propose encoding submitted problems as SAT 
(Boolean Satisifiability Instances).
Their model proposes that miners can select the traditional Hashcash PoW or a submitted 
${\cal NP}$-complete problem instance of the SAT problem.
They discusses how many `useful problem instances' papers or applications may not, in fact, be
useful in the immediate term.
They indicate, of course accumulating knowledge about problem solutions, even if not immediately applicable, may be
useful.
However, they propose industrially necessary ${\cal NP}$-complete problems that are useful in the ``immediate
practical value.''

\subsection{Technologies used}

Our proposal focuses on off-the-shelf technologies.
Some of the distributed computing systems such as Gridcoin, Golem and iExec leverage 
several of these technologies as well~\cite{Gridcoin,golem,iExec}.

\subsubsection{Docker}
Docker \cite{dockerio} is the basis of {\em Docker.com} and Docker is open source.
It is the most popular system for managing and executing virual containers.
Containers are OS-level  virtual machines.
That is, each container shares some of the underlying operating system's infrastructure.
This makes the containers more lightweight than full virtual machines.
At the same time, containers are not independent from their underying systems.
Namespaces isolate containers from their hosts~\cite{dockerlinuxjournal}. 

\begin{figure}
  \centering
  \includegraphics[scale=0.67]{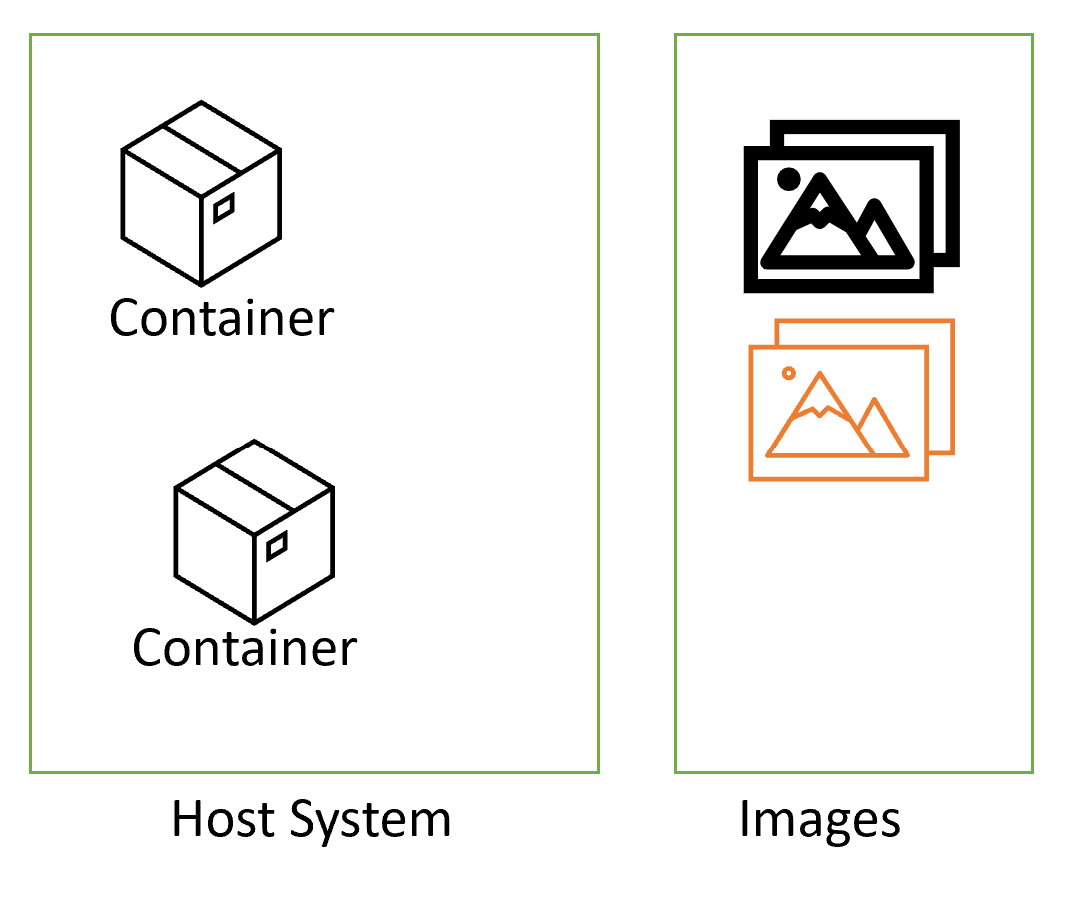}
  \caption{Basic containers}
  \label{fig:containers}
\end{figure}

Figure~\ref{fig:containers} shows the general structure of Docker containers.
Particularly, the plethora of images docker users maintain allows
very general computational problems to be executed.
These images and their standard installations can be used as the configuration set ups needed
for general PoVM computation.
Particularly, a dockerfiles can be a specific configuration.
Dockerfiles form the Docker repository can be validated using hash functions.

\subsubsection{Kubernetes}
 Kubernetes (K8s) \cite{kubernetes} is a Cloud Native Computing Foundation open source project.
K8s manages or orchestrates containerized services and workloads. 
 It provides a framework for running distributed systems resiliently.  K8s has many features including:
 Service discovery, load balancing, load management,
 automatic rollouts and rollbacks, self-healing, and configuration management.
When working with containers,  K8s works with container technologies such as Docker or LXC. 
Docker is the most used technology with K8s.

Kubernetes resource files are a good start towards orchestration SLAs.
These resource files describe how K8s should balance loads, structure docker instances, and scale.
Another type of SLA contract or file may be best for enforcing instructions-per-second, flops, etc.

\subsubsection{Skupper}
\begin{figure}
  \centering
  \includegraphics[scale=0.6]{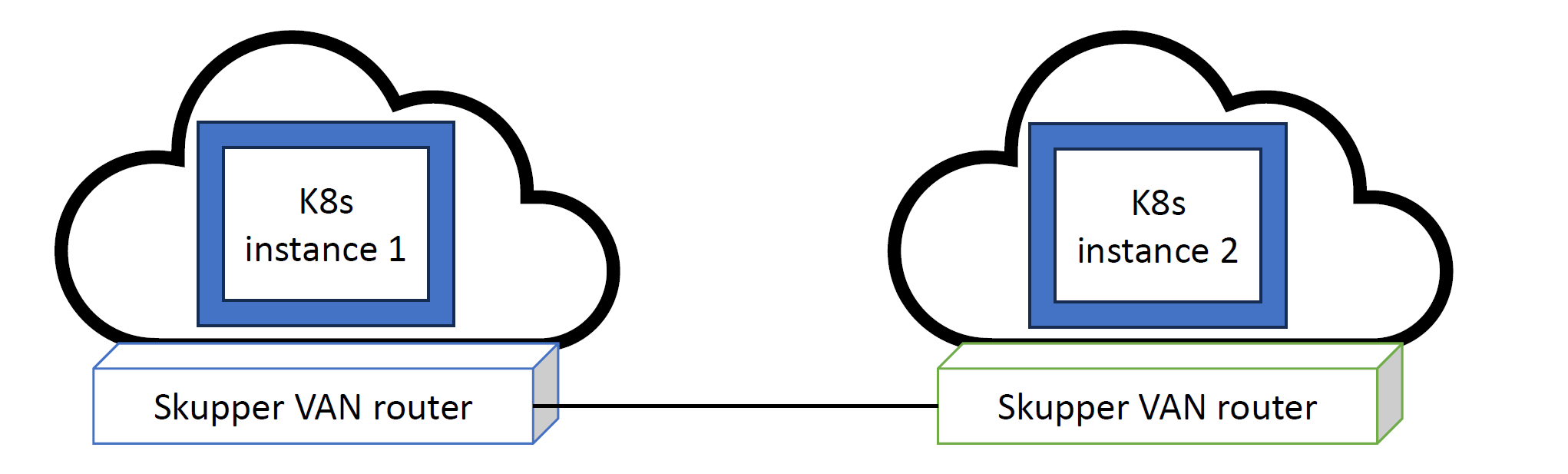}
  \caption{Skupper architecture showing Virtual Application Routers (VANs)}
  \label{fig:Skupper-van-routers}
\end{figure}

Skupper \cite{Skupperio}, is an Apache Software Foundation open-source system.
It creates a virtual application network (VAN) 
for multiple instances of K8s~\cite{Skupperio}. 
As shown in Figure~\ref{fig:Skupper-van-routers}, Skupper connects multiple K8s instances using virtual application routers.
Each K8s instance contains its own networks and containers.
Skupper works via K8s namespaces which separate services within a cluster and allows for managed instances of K8s.

Skupper is not a multiparty concensus based system.
So, it does not fit the general requirements for a blockchain-like system.
Furthermore, Skupper's work is not documented in a blockchain by PoW systems.

\section{System structure}

Golem, Gridcoin, iExec and SONM all use a number of methods to ensure the validity of their outsourced and off-chain 
computation~\cite{golem,Gridcoin,iExec,SONM}.
A key method these systems use for validation of outsourced and off-chain job computation is redundantly 
computing jobs. 
Then they compare the outcomes of these computations.
Another important aspect they have is to use reputation scoring.
That is, if a computation supplier does not return computed values as expected, then
they will earn a poor reputation.
Job expectations are based on concensus.
In addition, following the systems just mentioned,
if compute suppliers with poor reputation scores still offer their services, these services may 
be paid less due to their reputations.

\subsection{Technology stack}
This section gives background on our tech stack. 
The three primary technologies being used are Docker,  K8s, and Skupper. 
Each of these systems runs on container-based systems.
These container-based systems share aspects of the underlying host.
Sharing infrastructure may also lead to security issues.

\subsection{Docker}

This project uses Docker to deploy containers. 
Many of these containers are PoVMs.  
Other containers can be used to run Skupper, K8s or other systems.
The implementation also uses the Docker API to call on the specific endpoints enabling the creation, 
deletion, and monitoring of initiated containers. 
Customer submitted computations are executed by these containers. 

\subsection{Kubernetes}

K8s deploys and manages the containers for performing proof-of-VM rather than PoW mining 
computations. 
Particularly, PoVM is used in lieu of hashcash PoW. 
PoVM is also leveraged to create all the necessary components for Skupper
so it can connect containers via K8s instances across the internet for sharing resources.
These K8s instances perform self-healing 
on containers, elastic scaling, and general orchestration.

\subsection{Skupper}
Skupper connects  K8s clusters to manage and share resources.
That is, each K8s cluster has containers that each substitute for a `miner.' 
Each K8s instance contributes their containers for specific period of time. 
In any case, measuring the amount of computational resources is important.
Just measuring instructions per second will not be enough.
Each container running jobs for, say $T=24$-hours, to 
reduce the \cooo \ footprint of a typical blockchain system. 
One block may be expected to be computed every $m$ minutes or so by overlapping container start times.
If, for example, the period of computation-time is 24 hours, then every $m$ minutes new containers can get jobs from the
job queue and process them for 24 hours.
This may give more percision for block computation.

We must ensure each job is correctly computed. 
That is, there is no accidental or purposeful deception in any of the computation.

For example, some jobs can be computed by independent containers from different K8s clusters.
Their solutions can be compared to validate the correctness or incorrectness of the calculations.
This was implemented in Golem~\cite{golem} among other systems.
This can be easily modeled using probability theory. 
Then trade-offs can be made with the total cost and \cooo \ footprints as compared to 
repeated calculations.

\subsection{Technical gaps}

Our proof of technology does not include certain technical features that 
make it conform to be a single blockchain system.

\begin{enumerate}

\item We have not implemented a multiparty concensus-based lottery using
	multiparty generated randomness.  This can be done using smart contracts.

\item We do not have a multiparty concensus-based job queue. This can be done using smart contracts.

\item We do not have a solution for running Skupper by multiparty concensus. This may require a large amount of work.
	It seems to be quite challenging to adapt Skupper to be run as a multiparty concensus system.

\item Kubernetes is not run by multiparty concensus.
	It seems it would be very challenging and unwise to adapt K8s to be run as a multiparty concensus system nor does 
	K8s seem amenable to such modifications.
	
\item PoVM - multiparty concensus-based validation of job computation. 
The validations will require work with formalizing and validating configurations and SLAs using multiparty concensus.

\end{enumerate}

\section{Implementation}

This section gives more details about our solution. 
It starts with a basic setup of the technologies.
Next, it goes into a detailed discussion of how our system works.
We have a basic proof of technology demonstration that can be run.

\subsection{Our Solution}

Docker, K8s, and Skupper are set up so customers submit jobs to peer distributed network
though a webpage.
These jobs are put on the job queue.
Currently the job queue is not managed by multiparty PoW concensus.
However, it is possible to have a smart contract manage the job queue.

Transactions for blockchain processing can be put in a mempool as in standard blockchains.
Each host supplies its own instance of a K8s cluster consisting of 
Docker containers that are listening for work to compute. 
These K8s instances are run from Skupper.
The K8s clusters operate in their own separate K8s namespaces. 
This logically separates the clusters. 
Each host itself is part of a private network connected to many other hosts 
through the Skupper network. 

\begin{figure}
  \centering
  \includegraphics[scale=0.7]{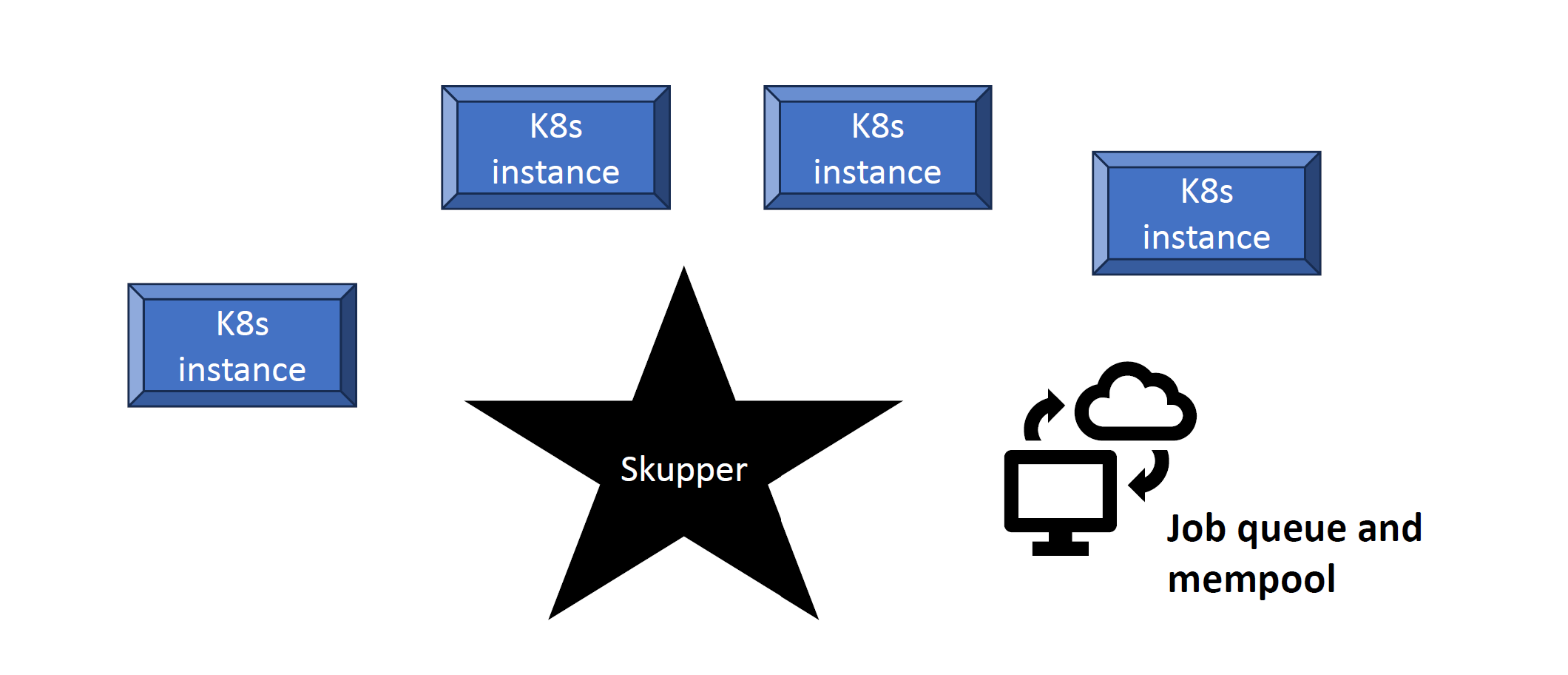}
  \caption{Low C02 blockchain architecture}
  \label{fig:our-architecture}
\end{figure}

Figure~\ref{fig:our-architecture} shows our basic centeralized architecture.
Here Skupper joins the K8s instances.  
These instances are run by Skupper.
The frontend allows jobs to be submitted to the job queue.
Then Skupper distributes jobs to K8s instances.
These K8s instances orchestrate Docker containers to run the submitted jobs.

A useful goal is to find a substitute for Skupper.
This substitute will run using multiparty concensus.

The demonstration that we run works with calculations on generating successive coin flips. 
Each working container uses a pseudo random coin flipper to determine how many coin flips the container 
must do to get a string of $k$ heads.
This job is interesting based on its mean and variance.
In particular the expected number of coin flips to get a sequence of $k$ heads is $2^{k+1} -2$.

The system's goal is to find the lucky pseudo-random seed that gets a string of $k$ heads in a row fastest.

\begin{enumerate}
  \item Once the Docker instances start up, the backend and frontend listener K8s configurations are set up.
  \item By accessing the local frontend listener IP given by  K8s on a local browser, user can submit the number of heads to get 
  successively in a row.
  \item Once submitted, the frontend server local to the host queries the Skupper network and pick another online host’s  
  K8s namespace at random and send the packaged data to a specifically targeted port.
  \item The receiving host’s frontend listener picks up the packaged data and calls the internal backend listener 
  to spawn a container that will flip the coins.
  \item The container computes the total number of flips needed to get $k$ heads.
  \item The container packages the answer and returns it to the backend listener which then talks to the original 
  host through Skupper and provides the result.
  \item The initial host opens the packaged response and updates the site to reflect the result from the request.
\end{enumerate}



%
%
%
\subsection{Running the demo}

The prerequisites for showing our proof of technology are given here. 
We use Mac OS X to run Docker desktop and Skupper.
 The following software needs to be installed prior to following this quick start guide:
\begin{itemize}
  \item Docker (w/Docker Desktop)
  \item Kind (K8s local cluster manager for Docker nodes)
  \item If on a Mac, then use Docker-mac-net-connect (via Homebrew)
\end{itemize}

We use kind for local management of our K8s instances. 
The choice of local management of K8s was to make this demonstration easy and inexpensive to run.

Once all the pre-requisites have been installed, clone the following project repository into a folder of your choice
from the repo~\cite{source-code}.

After cloning the repository, the majority of the instructions should be on the documentation provided, however, 
for the sake of simplicity, the following sub-points help with the demo.

\begin{enumerate}
  \item Create the KIND cluster
  \item Create  K8s configuration files
  \item Set-Up the Metal Load Balancer
  \item Connect all local namespaces via Skupper
  \item Apply all  K8s configurations for each namespace
  \item Create all  K8s deployments and expose the relative ports
  \item Test the technology.
\end{enumerate}

Now a basic Skupper network consisting of a head-sequence flipper example 
is available. 
The only required piece needed is to connect to the internal IP of the frontend docker container. 
Once the page loads, 
entering a number for the input causes the system to flow and prompt a response from a 
container run in another namespace.

\section{Future directions}

Performing percise \cooo \ analysis.
Adding a multi-signature scheme such as a Schnorr signature~\cite{Schnorr} where needed.

Exploring an iterative redundancy algorithm due to Brun, 
Edwards, Bang, and Medvidovic~\cite{brun}.
See also~\cite{fault-tolerant}.
It is notable that Brun, et al.~\cite{brun} discuss testing with the volunteer science 
system BOINC~\cite{boinc}.
Brun, et al. show how to compare the results of several PoVMs performing the same
computation.
These redundant calculations help prevent fraudulent computation.
For instance a VM that either accidentally or purposely does not do the correct computation.
These redundant calculations have been implemented in many systems such as Golem~\cite{golem}.

Given a single job $j$ to compute, 
{\em  $k$-vote traditional redundancy (TR)}, where $k$ is an odd integer, is

\begin{enumerate}

\item
	Clone job $j$ onto $k$ different but identical machine instances
	
\item
	Run one job $j$ on each of $k$ identical but independent containers
	
\item	
	Assume all job outputs are Boolean.
	All jobs outputs are compared and the majority of votes wins.

\end{enumerate}

Since $k$ is odd, it is pointed out that at least, $\frac{k+1}{2}$
of the same answer is evidence of a valid computation.
Our objective is to ensure that $k$ is far smaller than the number of expected 
PoW nodes.

Let $T$ be the total time for computing each clone for one PoVM job.
In the worst-case we compute all $k$ jobs.
For comparison, the PoW costs $p$ and there are $w$ total miners computing each
of these hashcash functions.
Thus, $\tau$ compares the basic cost of PoVM.

\begin{eqnarray*}
\tau & = & \left( k T + c \right) - p w
\end{eqnarray*}

We can also leverage checkpointing to enhance validation of evidence for PoVMs.
The checkpoints can be computed on all cloned jobs that are redundantly computing the same work
for validation.
In our case, checkpoints or their sub-checkpoints can be computed at the same program events for all clones.
We can compare clone checkpoints to validate different jobs are doing the expected computation.
These checkpoints may be kept secret by message-digesting them for each clone.
These comparisons be done using message-digest hash functions.
There are challanges to this approach.
For example, in asynchronous or randomized algorithms, different intermediate states may differ on different machines,
 yet these systems may compute the same final result.
Such cases require careful assessment of the checkpointing.

All of our technical gaps are of interest in the future.
That is, making the entire system run using multiparty concensus.
Perhaps the hardest part here is replacing Skupper.

\section{Conclusion}
Blockchains that use hashcash PoW consume a lot of energy. 
The proof of tech solution given here provides an alternative approach to that utilizes a peer to peer 
based blockchain network. 
Our approach employs containers as PoVM. This gives a way to condense computations along with a succinct
 method using container 
orchestration. 
The tech stack for the Low Carbon Blockchain Proof of Tech includes Docker,  K8s, and Skupper. 
Given the simplicity of the approach, there are many ways this solution can be extended and built on. 

There are a number of improvements that can be made to give a true multiparty concensus system.

\section{Thanks}

Dmitri Udler, Nadia Udler and
Alex Kachergis helped with a number of interesting discussions.
We also express our appreciation and gratitude to Adrian Torres and Ghulam Afzal.
Thanks to Roger Benites for his help and insight.

\vspace{2cm}

\section*{Authors}

\noindent {\bf Agron Gemajli} earned his BS degree in Computer Science from the University of Connecticut.
He is currently a graduate student studying for this masters in cybersecurity at New York University.
Agron is currently working as a cybersecurity analyst for Synchrony Financial. \\

\noindent {\bf Shivam Patel} earned his BS degree in Computer Science from the University of Connecticut.
 He is currently a graduate student studying for his Masters in Management of Technology at New York University.
Shivam works as a client engagement and technical project manager for Synchrony Financial. \\

\noindent {\bf Phillip G. Bradford} earned his BA degree in mathematics from Rutgers University, his MS in
computer science from the University of Kansas, and his PhD in computer science from Indiana University.
He was a post-doctoral fellow at the Max-Planck-Institut f\"ur Informatik.
He has worked for several prominent firms such as Blackrock and GE as well as with a number of startups.\\

\end{document}